\documentclass[prl,aps,twocolumn,amsmath,amssymb]{revtex4}
\usepackage[english]{babel}
\usepackage{natbib}
\usepackage{amsfonts}
\usepackage{amssymb}
\usepackage{graphicx}%
\usepackage{dcolumn}
\usepackage{amsmath}
\usepackage[ansinew]{inputenc}
\usepackage{psfig}
\usepackage{amssymb}

\begin{document}
\date{\today}
 \title
{Response to tilted magnetic fields in $Bi_2Sr_2CaCu_2O_8$ with columnar defects : 
Evidences for transverse Meissner effect}
\author{V. Ta Phuoc, E. Olive, R. De Sousa, A. Ruyter, L. Ammor and J.C. Soret
}
\affiliation{LEMA, CNRS- FRE 2077, Universit\'e Fran\c{c}ois Rabelais, 37200 Tours, France
}


\begin{abstract}
	The transverse Meissner effect (TME) in the highly layered 
superconductor $Bi_2Sr_2CaCu_2O_{8+y}$ with 
	columnar defects is investigated by transport measurements. 
We present detailed evidence for the 
	persistence of the Bose glass phase for $H_{\perp}<H_{\perp c}$ : (i) 
the variable-range vortex hopping process for low currents crosses over to the 
half-loops regime for high currents; (ii) in both regimes near 
$H_{\perp c}$  the energy 
	barriers vanish linearly with $H_{\perp}$ ; (iii) the transition 
temperature is governed by 
	$T_{BG}(H_{\parallel},0) -T_{BG}(H_{\parallel},H_{\perp}) \sim |H_{\perp}| ^{1/\nu_{\perp}}$ 
with  $\nu_{\perp}=1.0 \pm 0.1$. Furthermore, 
above the transition as $H_{\perp}\rightarrow H_{\perp c}^+$, 
moving kink chains consistent with a commensurate-incommensurate transition scenario 
are observed. These results thereby clearly show 
	the existence of the TME  for $H_{\perp}<H_{\perp c}$ .
 \end{abstract}

\maketitle

  
  The interplay between elasticity,  interactions, thermal fluctuations 
and dimensionality of magnetic flux lines subject to pinning, yields a 
large variety of vortex 
phases in high-T$_c$ superconductors \cite{1}. The nature of these 
phases and how they depend on the pinning is far from being completely 
elucidated. 
The theory of pinning by correlated disorder, such as twin planes or 
amorphous columnar tracks created by heavy ions, has 
	been considered by Nelson and Vinokur (NV) \cite{2} and 
Hwa {\em et~al} \cite{3}. 
	In the case of parallel tracks, NV have shown that if the applied 
magnetic field {\bf H} is aligned with the columnar 
	defects, the low temperature physics of vortices is similar to 
that of the Bose glass (BG) \cite{4}, with the flux lines strongly 
	localized in the tracks leading to zero creep resistivity. When 
{\bf H} is tilted at an angle $\theta$ away from the column 
	direction, the BG phase with perfect alignment of the internal 
flux density {\bf B} parallel to the columns is predicted 
	to be stable up to a critical transverse field $H_{\perp c}$ 
producing the so-called transverse Meissner effect (TME) 
	\cite{2,5}. For $\theta >  \theta_c \equiv \arctan(\mu_0H_{\perp c}/B)$, 
the linear resistivity is finite resulting from 
	the appearance of kink chains along the transverse direction as 
discussed in Ref.\ \onlinecite{6}. Finally, above a still 
	larger angle, the kinked vortex structures disappear and {\bf B} 
becomes collinear with {\bf H}. Similar scenarios apply to vortex pinning by the twin 
boundaries \cite{2} and by the layered structure of the coumpound 
	itself \cite{7}.\\
		\indent Recently, the TME in untwinned single crystals of 
$YBa_2Cu_3O_7$ (YBCO) with columnar defects, has been 
	observed using Hall sensors \cite{8}. Previous magnetic 
measurements in anisotropic 
3D superconductors, such as YBCO, gave support to the 
presence of vortex lock-in phenomena, due to pinning 
by the twin boundaries \cite{9} or by the interlayers between 
the Cu-O planes \cite{10}. In the case of highly layered 
superconductors, such as $Bi_2Sr_2CaCu_2O_{8+y}$ 
(BSCCO), the lock-in transition was observed for {\bf H} tilted 
away from the layers \cite{11}, but the existence of the TME 
due to the pinning by columnar defects remained an open question.\\
	\indent In this letter, we present measurements of the electrical 
properties near the glass-liquid transition in BSCCO 
single crystals with parallel columnar defects, as a function of $T$ 
and $\theta$ for filling factors $f\equiv \mu_0H_{\parallel}/B_{\Phi}<1$. 
Below a critical angle $\theta_c(T)$, we observe a vanishing resistivity 
$\rho (J)$ for low currents. A detailed {\it quantitative} analysis of  
$\rho (J)$ indicates that the creep proceeds via variable-range 
vortex hopping 
(VRH) at low currents due to some disorder \cite{12}, crossing 
over to the half-loop (HL) regime at high currents. 
For $\theta > \theta_c(T) $, the very signature of a kinked vortex 
structure, consistent with a commensurate-incommensurate 
(CIC) transition scenario in (1+1) dimensions \cite{6}, is deduced 
from the critical behavior of the 
linear resistivity. All these results clearly demonstrate that when 
{\bf H} is tilted at $\theta < \theta_c$ away 
from the defects, the flux lines remain localized on columnar defects, 
and hence, the BG exhibits a TME.\\
	\indent The BSSCO single crystal was grown by a self-flux technique, 
as described elsewhere \cite{13}. The crystal of 
$1\times 1\times 0.030\ mm^3$ size with the c-axis along the shortest 
dimension has a $T_c$ of $89\ K$, a transition 
width of $\sim 1\ K$, and was irradiated along its c-axis with 
$5.8\ GeV$ Pb ions to a dose corresponding to 
$B_{\Phi}=1.5\ T$ at the $GANIL$ (France). Isothermal $I-V$ 
curves were recorded using a $dc$ four-probe method with a sensitivity of 
$\sim 10^{-10}\ V$ and a 
temperature stability better than $5\ mK$. {\bf H} was aligned 
with the tracks using the well-known dip feature 
occurring in dissipation process for $\theta =0^{\circ}$, and was 
tilted with an angular resolution better than $0.1^{\circ}$ away from the 
column direction for $f$ fixed.\\
	\indent Fig.\ref{fig1} shows a typical log-log plot of $V/I-I$ 
curves obtained varying $\theta$ for $f=1/3$ and $T$ fixed below the BG transition 
temperature determined at $\theta =0^{\circ }$ \cite{12}.
We observe a well-defined angular crossover at an angle $\theta_c$. 
For  $\theta< \theta_c$, data suggest that the BG phase persists. 
On the contrary, 
the existence of an 
Ohmic regime for $\theta> \theta_c$ indicates a liquid vortex-like state.\\ 
	\indent We first focus on the angular range $\theta< \theta_c$, 
and we consider the scenario where the BG phase is stable. Thus, 
one expects that 
excitations of some localized 
vortex lines lead to a nonlinear resistivity given by \cite{2}:
   \begin{eqnarray}  
\rho(J)=\rho_0\ exp\big(-\tilde E_K(J_c/J)^{\mu}/k_BT\big)
   \end{eqnarray}
where $\rho_0$  is a characteristic flux-flow resistivity and 
$\tilde E_K(J_c/J)^{\mu}$ represents the barriers 
against vortex motion. This expression is predicted to hold for 
various regimes of different behavior as the current probes 
different length scales in the BG. 
$\tilde E_K$ acts as a scaling parameter for the pinning energy 
and $J_c$ is the characteristic current scale of the creep process. 
When the current is large enough that the 
growth of vortex-loops excitations of a line from its pinning track 
does not reach out the neighboring 
tracks, the HL excitations are relevant and yield an exponent 
$\mu=1$ and $J_c\equiv J_1=U/(\Phi_0d)$, 
where $U$ is the mean pinning potential and $d=\sqrt{\Phi _0/B_{\Phi }}$ 
is the mean spacing between pins. With decreasing current, size of half-loops 
increases with the result that some disorder in the pinning potential 
becomes relevant. This 
situation yields the VRH process characterized by $\mu=1/3$ 
and by another important current scale 
$J_c\equiv J_0=1/(\Phi _0g(\tilde\mu)d^3)$, where $g(\tilde\mu)$  
denotes the density of pinning energies at the 
chemical potential of the vortex system. One expects that both 
$J_1$ and $J_0$ are insensitive to $H_{\perp}$. 
We therefore consider that the only important effect of $H_{\perp}$ 
is to lower $\tilde E_K$, according to the formula :
   \begin{eqnarray}  
\tilde E_K=E_K-\epsilon ^2\Phi _0dH_{\perp}
   \end{eqnarray} 
where $E_K$ is the mean kink energy, and the energy gain due 
to the tilt is obtained from the isotropic result 
$\Phi _0dH_{\perp}$  by applying the scaling rule 
[Eq.\ (3.12)] of the Ref.\ \onlinecite{1}.\\ 
	\indent We now present the following two-step method of 
fitting Eqs.\ (1) and (2) to the data, which allows to get a good 
understanding of the physics in our experiment. First, we plot in 
Fig.\ref{fig2} the natural logarithm of V/I versus 
$\tan\theta$  for $I$ fixed. Note that $\tan\theta$ is here directly 
proportional to $H_{\perp}$ since $f$ 
is kept constant. A linear variation $\ln(V/I)=M+N \tan\theta$ 
(expression designated 
below as $E1$) with $M$ and $N$ being current-dependent 
parameters is found for $\theta$ varying up to $\theta_c$ 
(solid lines), whereas above $\theta_c$ data deviate from this 
behavior (dotted lines). This finding is another argument 
supporting a true angular transition at $\theta_c$, as suggested 
in Fig.\ref{fig1}. It should be noted that we draw such 
a conclusion from the observation of two {\it independent} 
regimes of {\it different} behavior. One leads to a vanishing 
linear resistivity as $\theta\rightarrow\theta_c^+$, and the 
other is evidenced by probing the regions of 
nonlinear resistivity above and below $\theta_c$. Another 
result is that $N\sim I^{-\mu }$ with $\mu$ undergoing a 
jump from $1/3$ to $1$ at 
$I_2\approx 60\ mA$. Such a current crossover is clearly 
visible in  insert a of Fig.\ref{fig2} (filled symbols) where we 
plot together $N$ 
normalized by $I_0^{1/3}$ and $N$ normalized by $I_1$ versus 
$I$ depending upon whether $I$ is $<I_2$ or $>I_2$, respectively. 
Here, $I_0$ and $I_1$ are two currents evaluated 
on the basis of the BG model, as will be seen below. 
It therefrom follows that $N_{1/3}(I)=K(I_0/I)^{1/3}$ and $N_{1}(I)=KI_1/I$ 
where the labels $1/3$ and $1$ differentiate between two 
fits below and above $I_2$, respectively. In both these 
equations, $K\approx 0.6$ is a dimensionless constant. 
It should be noted that such a crossover at $I_2$ is clearly 
observed from $V/I-I$ curves as well (see Fig.\ref{fig1}), 
where a sudden increase of $V/I$ suggesting an increasing 
vortex motion, occurs at $I_2$. Second, we plot in Fig.\ref{fig3} 
and its insert the natural logarithm of $V/I$ versus $I^{-1/3}$ 
for $I<I_2$ and versus $I^{-1}$ for $I>I_2$, respectively. 
We verify that the expression $ln(V/I)=P-Q_{\mu}I^{-\mu}$ 
(expression designated below as $E2$) fits very well our 
experimental data (solid lines) with an exponent $\mu=1/3$ for 
$I<I_2$ and with $\mu=1$ for $I>I_2$. In $E2$, $Q_{\mu}$ 
is labeled using the above convention. The result is that $P$ 
is a constant (within the experimental errors), 
while $Q_{\mu}$ depends on $\theta$. In Fig.\ref{fig4} 
and its lower insert, we show $Q_{1/3}$ and $Q_{1}$ versus $\tan{\theta}$, 
respectively. A linear variation in $\tan{\theta}$ is observed 
consistently with expression $E1$. In upper insert of 
Fig.\ref{fig4}, we plot together $Q_{1/3}/I_0^{1/3}$ and 
$Q_{1}/I_1$ versus $\tan{\theta}$. 
The result is that the data are superimposable onto a single 
straight line with slope $\approx-0.6$ consistent with 
the value of $-K$, as may be verified by identifying the expression 
$E1$ with the expression $E2$.\\
	\indent In conclusion, $V/I=R_0\exp{\left[-\left(K'-K\tan{\theta}\right)\left(I_0/I\right)^{1/3}\right]}$
is clearly observed in our experiment for $I< I_2$ (solid lines in Fig.\ref{fig1}) while 
$V/I=R_0\exp{\left[-\left(K'-K \tan{\theta}\right)\left(I_1/I\right)\right]}$ is valid for data above 
$I_2$ (dashed lines in Fig.\ref{fig1}). In both fitting expressions 
$K$ and $K'$ depend on $f$ and $T$. In the case of data 
presented in Fig.\ref{fig1}, 
$K\approx 0.6$ and $K'\approx 0.4$; furthermore 
$I_0=15\ A$ and $I_1=0.4\ A$ are evaluated from 
BG model (see below), and 
$R_0\approx 1.1\ \mu\Omega$ is five orders of magnitude 
lower than the normal resistance. 
Therefore, for $\theta<\theta_c$ we have rather strong 
evidence of two separate (albeit related) vortex creep processes 
peculiar to a BG : the HL expansion with $\mu=1$ which 
is cut off by the crossover at $I_2$ into the VRH process 
with $\mu=1/3$. Moreover, we note that for each filling 
factor investigated in our experiment ($f=2/15,\ 1/3,\ 2/3$), we 
observe a {\it continuation} of the VRH process evidenced at 
$\theta=0^{\circ}$ \cite{12}. We therefore argue that the BG 
phase remains stable up to a critical tilting angle $\theta_c$, as 
predicted by NV \cite{2}.\\
	\indent To test the accuracy of the above view, we estimate 
from the theory \cite{1,2}, first the characteristic current 
scales, secondly the current crossover, and then the energy 
barriers in Eq.\ (1). $J_0$ only depends on 
$g(\tilde{\mu})$, the density of pinning energies evaluated at 
the chemical potential of the vortex system. Although a form of $g(x)$ is not yet available, 
an estimate of $g(\tilde{\mu})$ can be done in terms 
of the bandwidth of pinning energies $\gamma$, due to the disorder, and hence 
$g(\tilde{\mu})\approx1/\left(d^2\gamma\right)$ so that 
$J_0\approx \gamma /(\phi _0d)$. Following Blatter {\em et~al} \cite{1}, 
$\gamma=t_d+\gamma_i$ where 
$t_d\approx U/\sqrt{E_K/k_BT}\exp\left( -\sqrt{2}E_K/k_BT\right)$ 
estimates the dispersion in the 
pinning energies resulting from disorder in the position of 
tracks \cite{2} and $\gamma_i$ arises from some on-site randomness. We shall 
assume random defect radii as the main source of on-site disorder. 
A reasonable estimate of $\gamma_i$ is then given by the 
width of the distribution of pinning energies $\tilde{P}(U_K)=P(c_K)dc_K/dU_K$ 
where $U_K$ is the binding energy of a defect with radius 
$c_K$ and $P(c_K)$ is the probability distribution of the 
defect radii. A realistic 
$pdf$ is a normal law centered at $c_0=45$ \AA \ with a 
standard deviation of $6$ \AA, as shown in figure 1 of 
Ref.\ \onlinecite{12}. 
Thus, using the formula 
$U_K=(\phi _0^{2}/8\pi \mu _0\lambda _{ab}^{2})ln[1+(c_K/\sqrt{2}\xi _{ab})^2]$ 
\cite{1,2} where 
$\lambda _{ab}$ and $\xi_{ab}$ are respectively the planar 
penetration depth and the planar coherence length, we determine $U_0=<U_K>$ 
representing the energy scale for the vortex pinning. Another 
important energy scale corresponding to vortex positional fluctuations is 
$T^{\star}=(k_Bc_0/4\xi _0G_i)\sqrt{ln(a_0/\xi _{ab})}\ (T_c-T)$ 
where $G_i$ is the 2D Ginzburg number and $a_0=\sqrt{\phi _0/B}$ 
is the vortex-lattice constant. $U_0$ and $T^{\star}$ completely 
characterize the system of vortices and columnar defects. Thus, it can be inferred  
$E_K=(d/\sqrt{2}\xi _{ab})\ T^{\star} \sqrt{f(T/T^{\star})}$ 
where $f(x)=x^2/2\ exp(-2x^2)$ accounts for entropic effects, and  
$U=U_0\ f(T/T^{\star})$. Using appropriate parameters for BSCCO 
($\lambda_0\approx 1850\AA$, $\xi_0\approx20\AA$ and $G_i\approx0.2$) 
we estimate $J_0\approx 10^9\ A/m^2$ and $J_1\approx 2.6\ 10^7\ A/m^2$ 
for $f=1/3$  and $T=68\ K$. The balance between the barrier 
energy for the VRH process $U_{VRH}=\tilde{E_K}(J_0/J)^{1/3}$ 
and the barrier for the HL regime $U_{HL}=\tilde{E_K}\ J_1/J$ 
determines the crossover current $J_2=(J_1/J_0)^{1/2}J_1=U^{3/2}/(\phi _0d\sqrt{\gamma})$. 
Thus, we have $J_2\ll J_1\ll J_0$ so that the 
crossover between the VRH process and the HL regime takes place at $J_2$. 
Assuming uniform currents into the sample, we find $I_2\approx 60\ mA$ 
in excellent agreement with the experiment (Fig.\ref{fig1}). A more 
complete quantitative evaluation of the consistency of the data with the model is 
obtained for different values of $f$ and $T$, as well as for different 
values of $d$ (insert in Fig.\ref{fig2}). In particular, we find that the effect of $d$ 
on $I_2$ is in good agreement with the prediction $I_2 \propto 1/d$. 
Finally, evaluating $U_{VRH}$ and $U_{HL}$ with $\tilde{E_K}$ from Eq.\ (2), 
we obtain using again the above usual parameters of BSCCO with 
$\epsilon \approx 1/200$ that $E_K/k_BT\approx 0.3$ and 
$\varepsilon ^2\phi _0dH_\bot /k_BT\approx 0.7\tan\theta $ for 
$f=1/3$ and $T=68\ K$, which correspond to the values of our fitting 
parameters $K'$ and $K $ (see upper insert in Fig.\ref{fig4}). 
Thus, the experiment is also in good agreement with the barriers $U_{VRH}$ 
and $U_{HL}$ theoretically predicted below and above $I_2$, respectively.\\        
	\indent Finally, we pass on to the linear regime observed above 
$\theta_c$ (see Fig.\ 1). For $\theta> \theta_c$, 
Hwa {\em et~al} \cite{6} predict on the basis of the CIC transition, 
the appearance of free moving chains of kinks oriented 
in the $H_{\perp}$ direction leading to a critical behavior of the linear resistivity 
$\rho\sim(\tan{\theta}-\tan{\theta_c})^{\upsilon }$ characterized by an 
exponent $\upsilon =1/2$ $[\upsilon =3/2]$ in $(1+1)$ $[(2+1)]$ 
dimensions. This type of behavior with $\upsilon =1/2$ is evident in Fig.\ref{fig5} 
which shows a plot of $R/R_c$ versus $\tan{\theta}-\tan{\theta_c}$ 
for different values of $f$ (or equivalently $H_{\parallel}$) and  
$T<T_{BG}(H_{\parallel},0)$. Here, $R_c$ is a scaling parameter 
comparable to $R_0$, and for  $f$ fixed, $\theta_c(T)$ is determined fitting 
$R=R_c(\tan{\theta}-\tan{\theta_c})^{1/2}$ to the data. In Fig.\ref{fig2}, 
the arrow indicates $\theta_c$ obtained in this way. The insert of the figure 5 shows 
$t(H_{\parallel},H_{\perp})\equiv (T_{BG}(H_{\parallel},0)-T_{BG}(H_{\parallel},H_{\perp})/T_{BG}(H_{\parallel},0)$ 
as a function of $H_{\perp}$ 
where $T_{BG}(H_{\parallel},H_{\perp})$ is obtained through 
the inversion of $\theta_c$ and $T$. It turns out that $t(H_{\parallel},H_{\perp})$ is 
independent of $H_{\parallel}$. The solid line shown in insert is a fit to the data  
following $t(H_{\parallel},H_{\perp})\sim |H_{\perp}|^{1/\nu_{\perp}}$ 
with $\nu_{\perp}=1.0\pm 0.1$, as recently 
suggested from numerical simulations \cite{14} and observed in 
$(K,Ba)BiO_3$ \cite{15}.Moreover, such a value of $\nu_{\perp}$ is excellently 
consistent with the result previously found using the scaling theory of the 
BG transition at $H_{\perp}=0$ \cite{12} .\\
	\indent In summary, we demonstrate from transport measurements 
the stability of the BG phase in BSCCO single crystal with columnar defects, when 
{\bf H} is tilted at $\theta \leq \theta _c$ away from the column direction. 
We explain the critical behavior of the linear resistivity on the basis of the CIC 
transition of kink chains right above $\theta _c$, as predicted in 
$(1+1)$ dimensions for one-dimensional correlated disorder. 
This implies that the appearance 
of the TME is concomitant with the vanishing of the linear resistivity 
at $\theta _c$. As a consequence, our results support the scenario 
for an usual BG to 
liquid transition in irradiated BSCCO in contrast to a puzzling 
two-stage BG to liquid transition recently observed in untwinned YBCO with columnar 
defects \cite{8}.



\begin{figure}
\caption{\label{figure1}$V/I-I$ curves for tilted magnetic fields. From the right to the left 
$\theta =0$, $5$, $10$, $15$, $20$, $25$, $30$, $35$ and $40^\circ$. 
The curved lines are a fit of the Bose glass theory to the data.} 

\end{figure}

\begin{figure}
\caption{\label{fig2}
Angular dependence of $ln(V/I)$ for different $I$. The solid 
lines are a fit of $ln(V/I)=M+N\tan\theta $ to the data; the dotted lines are a guide 
for the eye. Insert: $N/I_0^{1/3}$ vs $I$ (left axis) and $N/I_1$ vs 
$I$ (right axis) with $I_0$ and $I_1$ determined from theory. $(a)$ $f=1/3$ and 
$T=68.553K$ (filled symbols), $f=2/15$ and $T=69.136K$ 
(open symbols). $(b)$ $f=1/3$ and $T=59.960K$ in another crystal with $B_{\Phi}=0.75\ T$. 
The straight lines are fits whence we obtain $\mu =0.33\pm 0.05$ 
(solid line) for $I<I_2$ and $\mu =1.0\pm 0.1$ (dashed line) for $I>I_2$.}
\end{figure}

\begin{figure}
\caption{\label{fig3}
$ln(V/I)$ vs $I^{-1/3}$ for $I<I_2$ and $\theta <\theta _c$. 
Insert: $ln(V/I)$ vs $I^{-1}$ for $I>I_2$ and $\theta <\theta _c$. 
In both plots the lines are a fit to the data using the form $ln(V/I)=P-Q_\mu I^{-\mu }$ 
with $\mu =1/3$ or 1 according to whether $I$ is $<I_2$ or $>I_2$.}
\end{figure}

\begin{figure}
\caption{\label{fig4}
Angular dependence of $Q_{1/3}$ defined in the regime $I<I_2$ 
(see Fig.\ref{fig3}). Lower insert: angular dependence of $Q_1$ 
defined in the regime $I>I_2$. Upper insert: angular dependence of 
$Q_{1/3}/I_0^{1/3}$ and $Q_1/I_1$. In each plot the straight line is a least-square fit.}
\end{figure}

\begin{figure}
\caption{\label{fig5}
Critical behavior of the linear resistance for different values of $f$ and $T$.
The curved line is a fit to the data using 
$R=R_c(tan\theta -tan\theta _c)^{1/2}$ in accordance with a CIC transition scenario. 
Insert: angular dependence of the Bose glass transition. The solid line is a fit of 
$T_{BG}(H_{\parallel},0) -T_{BG}(H_{\parallel},H_{\perp}) \sim |H_{\perp}| ^{1/\nu_{\perp}}$
with $\nu _{\perp} =1.0\pm 0.1$ to the data, the point $(0,0)$ included.}
\end{figure}
  
\end{document}